\documentclass[10pt,onecolumn]{IEEEtran}
\usepackage{times,amssymb,amsmath,amsfonts,float,nicefrac,color,bbm,tikz-cd}
\usepackage{euscript,graphics}
\usepackage[dvips]{epsfig}
\usetikzlibrary{matrix,arrows,decorations.pathmorphing}

\newtheorem{theorem}{Theorem}[section]

\newtheorem{proposition}[theorem]{Proposition}

\newtheorem{definition}{Definition}
\newtheorem{remark}{Remark}[section]

\newtheorem{cnstr}{Construction}

\newtheorem{xmpl}{Example}

\newcommand\qed{\rule{2mm}{2.5mm}\medskip}

\newcommand{\remove}[1]{}
\renewcommand{\tilde}{\widetilde}

\newcommand\nd{\noindent}

\def\supp{\qopname\relax{no}{supp}}

\newcommand{\ceilenv}[1]{\left\lceil #1 \right\rceil}

\newcommand\nc\newcommand
\nc\bfa{{\boldsymbol a}}\nc\bfA{{\bf A}}\nc\cA{{\mathcal A}}
\nc\bfb{{\boldsymbol b}}\nc\bfB{{\bf B}}\nc\cB{{\mathcal B}}
\nc\bfc{{\boldsymbol c}}\nc\bfC{{\bf C}}\nc\cC{{\mathcal C}}
\nc\bfd{{\boldsymbol d}}\nc\bfD{{\bf D}}\nc\cD{{\mathcal D}}
\nc\bfe{{\boldsymbol e}}\nc\bfE{{\bf E}}\nc\cE{{\mathcal E}}
\nc\bff{{\boldsymbol f}}\nc\bfF{{\bf F}}\nc\cF{{\mathcal F}}
\nc\bfg{{\boldsymbol g}}\nc\bfG{{\bf G}}\nc\cG{{\mathcal G}}
\nc\bfh{{\boldsymbol h}}\nc\bfH{{\bf H}}\nc\cH{{\mathcal H}}
\nc\bfi{{\boldsymbol i}}\nc\bfI{{\bf I}}\nc\cI{{\mathcal I}}
\nc\bfj{{\boldsymbol j}}\nc\bfJ{{\bf J}}\nc\cJ{{\mathcal J}}
\nc\bfk{{\boldsymbol k}}\nc\bfK{{\bf K}}\nc\cK{{\mathcal K}}
\nc\bfl{{\boldsymbol l}}\nc\bfL{{\bf L}}\nc\cL{{\mathcal L}}
\nc\bfm{{\boldsymbol m}}\nc\bfM{{\bf M}}\nc\cM{{\mathcal M}}
\nc\bfn{{\boldsymbol n}}\nc\bfN{{\bf N}}\nc\cN{{\mathcal N}}
\nc\bfo{{\boldsymbol o}}\nc\bfO{{\bf O}}\nc\cO{{\mathcal O}}
\nc\bfp{{\boldsymbol p}}\nc\bfP{{\bf P}}\nc\cP{{\mathcal P}}
\nc\bfq{{\boldsymbol q}}\nc\bfQ{{\bf Q}}\nc\cQ{{\mathcal Q}}
\nc\bfr{{\boldsymbol r}}\nc\bfR{{\bf R}}\nc\cR{{\mathcal R}}
\nc\bfs{{\boldsymbol s}}\nc\bfS{{\bf S}}\nc\cS{{\mathcal S}}
\nc\bft{{\boldsymbol t}}\nc\bfT{{\bf T}}\nc\cT{{\mathcal T}}
\nc\bfu{{\boldsymbol u}}\nc\bfU{{\bf U}}\nc\cU{{\mathcal U}}
\nc\bfv{{\boldsymbol v}}\nc\bfV{{\bf V}}\nc\cV{{\mathcal V}}
\nc\bfw{{\boldsymbol w}}\nc\bfW{{\bf W}}\nc\cW{{\mathcal W}}
\nc\bfx{{\boldsymbol x}}\nc\bfX{{\bf X}}\nc\cX{{\mathcal X}}
\nc\bfy{{\boldsymbol y}}\nc\bfY{{\bf Y}}\nc\cY{{\mathcal Y}}
\nc\bfz{{\boldsymbol z}}\nc\bfZ{{\bf Z}}\nc\cZ{{\mathcal Z}}
\nc\od{{\bar d}}\nc\ow{{\bar w}}\nc\odelta{{\bar\delta}}
\nc\ox{{\bar x}}\nc\oy{{\bar y}}\nc\ou{{\bar u}}
\nc\oh{{\bar h}}

\newcommand\ff{{\mathbb F}}
\newcommand\kk{{\mathbbm k}}
\newcommand\pp{{\mathbb P}}

\nc\ellone{{\ell_1}}
\nc\elltwo{{\ell_2}}
\nc\ellinf{{{\ell_\infty}}}
\nc\ip[2]{\langle #1,#2\rangle}

\newcommand{\beeq}{\begin{eqnarray*}}
\newcommand{\eneq}{\end{eqnarray*}}

\begin{document}

\sloppy

\title{Locally recoverable codes on algebraic curves} 

\author{\IEEEauthorblockN{Alexander Barg$^{a}$}\quad
\and \IEEEauthorblockN{Itzhak Tamo$^{b}$}\quad
\and \IEEEauthorblockN{Serge Vl{\u a}du{\c t}$\,^{c}$}}
\maketitle
{\renewcommand{\thefootnote}{}\footnotetext{

\vspace{-.2in}
 
\noindent\rule{1.5in}{.4pt}

\nd $^{a}$ Dept. of ECE and ISR, University of Maryland, College Park, MD 20742 and IITP, Russian Academy of Sciences, Moscow, Russia. Email abarg@umd.edu. Research supported by NSF grants CCF1422955 and CCF1217245.

\nd $^{b}$ Dept. of EE-Systems, Tel Aviv University, Tel Aviv, Israel. Research done in part while at the Institute for Systems Research, University of Maryland, College Park, MD 20742. Email zactamo@gmail.com. Research supported in part by NSF grant CCF1217894.

\nd $^{c}$ Institut de Math{\'e}matiques de Marseille, Aix-Marseille Universit{\'e}, IML, 
Luminy case 907, 13288 Marseille, France, and IITP, Russian Academy of Sciences, Moscow, Russia. Email
serge.vladuts@univ-amu.fr.

}}
\renewcommand{\thefootnote}{\arabic{footnote}}
\setcounter{footnote}{0}
\thispagestyle{empty}

\vspace*{-.1in}
\begin{abstract}
A code  over a finite alphabet is called locally recoverable {(LRC code)} if every symbol in the encoding is a function of a small number (at most $r$) other symbols of the codeword.  In this paper we introduce a construction of LRC codes on algebraic curves, extending a recent construction of Reed-Solomon like codes with locality. 
We treat the following situations: local recovery of a single erasure, local recovery of multiple erasures, and
codes with several disjoint recovery sets for every coordinate (the {\em availability problem}).
For each of these three problems we describe a general construction of codes on curves and construct several families of LRC
codes. \textcolor{black}{We also describe a construction of codes with availability that relies on automorphism groups of curves.}

We also consider the asymptotic problem for the parameters of LRC codes on curves. We show that the codes obtained from
asymptotically maximal curves (for instance, Garcia-Stichtenoth towers) improve upon the asymptotic versions
of the Gilbert-Varshamov bound for LRC codes. 
\end{abstract}
\vspace*{-.1in}
\section{Introduction: LRC Codes}
The notion of locally recoverable, or LRC codes is motivated by applications of coding to increasing reliability and efficiency of distributed storage systems.
Following \cite{gopalan2011locality}, we say that a code $\cC\subset \ff_q^n$ is LRC with locality $r$ if
the value of every coordinate of the codeword can be found by accessing at most $r$ other coordinates of this codeword, i.e.,
one erasure in the codeword can be corrected in a local way.
Let us give a formal definition. 
\vspace*{.05in}
\begin{definition}[LRC codes]\label{def:LRC} A code $\cC\subset \ff_q^n$ is LRC with locality $r$ if for every $i\in [n]:=\{1,2,\dots,n\}$
there exists a subset $A_i\subset [n]\backslash \{i\}, |A_i|\le r$ and a function $\phi_i$ such that for every codeword $x\in\cC$ we have
   \begin{equation}\label{eq:def1}
   x_i=\phi_i(\{x_j,j\in A_i\}).
   \end{equation}
This definition can be also rephrased as follows. Given $a\in \ff_q,$ consider the sets of codewords
   \begin{equation*}
   \cC(i,a)=\{x\in \cC: x_i=a\},\quad i\in[n].
   \end{equation*}
    The code $\cC$ is said to have  {locality} $r$ if for every $i\in [n]$ there exists a subset $A_i\subset [n]\backslash i, |A_i|\le r$
    such that the restrictions of the sets $\cC(i,a)$ to
the coordinates in $A_i$ for different $a$ are disjoint:
 \begin{equation}\label{eq:def}
 \cC_{A_i}(i,a)\cap \cC_{A_i}(i,a')=\emptyset,\quad  a\ne a'.
 \end{equation}
We use the notation $(n,k,r)$ to refer to the parameters of an LRC code of length $n$, cardinality $q^k,$ and locality $r$.
\end{definition}

The concept of LRC codes can be extended in several ways. One generalization concerns correction of multiple 
erasures (local recovery of several coordinates) \cite{prakash2012optimal,kamath2012codes}.

\vspace*{.05in}
\begin{definition}[LRC codes for multiple erasures]\label{def:LRC-rho} A code $\cC\subset \ff_q^n$ of size $q^k$ is said to have the $(\rho,r)$ {\em locality property} (to be an $(n,k,r,\rho)$ LRC code) where $\rho\geq 2$, 
if each coordinate $i\in [n]$ is contained in a subset $J_i\subset [n]$ of size at most $r+\rho -1$ 
such that the restriction $\cC_{J_i}$ to the coordinates in $J_i$ forms a code of distance at least $\rho$. 
\end{definition}
\vspace*{.05in}
We note that for $\rho=2$ this definition reduces to Definition \ref{def:LRC}. Note also that
the values of any $\rho-1$ coordinates of $J_i$ are determined by the values of the  remaining $|J_i|-(\rho-1)\leq r$ coordinates, thus enabling local recovery.

Another extension of Definition \ref{def:LRC} concerns codes with {\em multiple recovery sets}; see, e.g., \cite{raw14,tam14b}. 

\vspace*{.05in}\begin{definition}[LRC codes with availability]\label{def:LRC-t} A code $\cC\subset \ff_q^n$ of size $q^k$ is said to have $t$ recovery sets (to be an LRC$(t)$ code) if
for every coordinate $i\in[n]$ and every $x\in \cC$ condition \eqref{eq:def1}
holds true for pairwise disjoint subsets $A_{i,j}\subset[n]\backslash\{i\},|A_{i,j}|=r_j,j=1,\dots,t.$ 
We use the notation $(n,k,\{r_1,r_2,\dots,r_t\})$ for
the parameters of an LRC$(t)$ code.
\end{definition}
\vspace*{.05in} Codes with several recovery set make the data in the
system better available for system users, therefore the property of several recovery sets is often called the {\em availability
problem}.

A variation of the above definitions, called {\em information locality}, assumes that local recovery is possible only for the message symbols of the codeword. For this reason, the codes defined above are also said to have {\em all-symbol locality property}. In
this paper we study only codes with all-symbol locality, calling them locally recoverable (LRC) codes.

Let us recall some of the known bounds on the parameters of LRC codes. 
				The minimum distance of an $(n,k,r,\rho)$ LRC code satisfies the inequality
				\cite{kamath2012codes}
    \begin{equation}\label{eq:rho}
    d\le n-k+1-\Big(\Big\lceil \frac kr\Big\rceil -1\Big)(\rho-1).
    \end{equation}
For the case of $\rho=2$ this result was previously derived in \cite{gopalan2011locality,pap12}
   			\begin{equation}
				d\leq n-k-\ceilenv{ \frac{k}{r}}+2.				 
				\label{eq:dist}
     		\end{equation} 
Below we call codes whose parameters attain these bounds with equality {\em optimal LRC codes.}	

The following bounds are known for the distance of codes with multiple recovery sets. Let $\cC$ be an $(n,k,\{r,\dots,r\})$ LRC$(t)$ code, i.e., an $(n,k)$ $q$-ary code with
$t$ disjoint recovery sets of size $r$, then its distance satisfies
   \begin{align}
   d&\le n-k+2-\Big\lceil\frac{t(k-1)+1}{t(r-1)+1}\Big\rceil \quad\textcolor{black}{\cite{raw14}},\cite{wang2014a}\label{eq:tbf-1}\\
   d&\le n-\sum_{i=0}^t\Big\lfloor\frac{k-1}{r^i}\Big\rfloor \quad\cite{TBF}. \label{eq:tbf}
   \end{align}
The bounds \eqref{eq:tbf-1}, \eqref{eq:tbf} obviously reduces to \eqref{eq:dist} for $t=1.$ 
\textcolor{black}{For $n\to\infty$ the bound \eqref{eq:tbf} is tighter than the bound \eqref{eq:tbf-1} for all $R=k/n,
0<R<1;$ see \cite{TBF} for more details.}
Generally, little is known about the tightness of these
bounds, even though \eqref{eq:tbf} can be shown to be tight for some examples of short codes \cite{TBF}.

The bounds \eqref{eq:dist}-\eqref{eq:rho} extend the classical Singleton bound of coding theory, which is attained by the 
well-known family of Reed-Solomon (RS) codes. The Singleton bound is obtained from \eqref{eq:dist} by taking $r=k,$
which is consistent with the fact that the locality of RS codes is $k.$ Therefore, if our goal is constructing codes of
dimension $k$ with small locality, then RS codes are far from being the best choice.
RS-like codes with the LRC property whose parameters meet the bound \eqref{eq:dist} for any locality value $r\ge 1$ were recently constructed in \cite{tam14a}.
Unlike some other known constructions, e.g., \cite{sil13,My-paper}, the codes in \cite{tam14a} are constructed over finite fields of cardinality comparable to the code length $n$ (the exact value of $q$ depends on 
the desired value of $r$ and other code parameters, but generally is only slightly greater than $n$). 

Similarly to the classical case of MDS codes, the length of the codes 
in this family is restricted by the size of the alphabet, i.e., in all the known cases $n\le q.$
The starting point of this work is the problem of constructing families of longer LRC codes. To address this problem, 
we follow the general ideas of classical coding theory \cite{TVN07}.
RS codes can be viewed as a special case of the general construction of geometric Goppa codes; in particular, good codes
are obtained from families of curves with many rational points. Motivated by this approach,
in this paper we take a similar view of the construction of the evaluation codes of \cite{tam14a}. We present
a general construction of LRC codes on algebraic curves for the 3 variants of the LRC problem defined above.

We begin with observing that the codes in \cite{tam14a} arise from a trivial covering map of 
projective lines, which suggests one to look at covering maps of algebraic curves. 
This results in a general construction of LRC codes on curves, and the codes obtained in this way turn out to have good parameters because many good curves (curves with many rational points) are obtained as covers of various kinds. 
Our construction is also flexible in the sense that it enables one to accommodate various restrictions arising from the locality property, for instance, constructing codes with small locality, or constructing $(n,k,r,\rho)$ LRC codes with local distance $\rho\ge 2.$ Similarly to \cite{tam14a}, in all the constructions local recovery of the erased coordinates can be performed by interpolating a univariate polynomial over the coordinates of the recovery set. 

As is well known, in the classical case some families of codes on curves have 
very good asymptotic parameters, and in particular, improve upon the asymptotic Gilbert-Varshamov (GV) bound that connects
the code rate and the relative distance. Here we show similar results for LRC codes on curves, improving upon the asymptotic
GV-type bounds for codes with a given locality $r$ as well as for $(n,k,r,\rho)$ codes for all $\rho\ge 2.$

The LRC Reed-Solomon codes in \cite{tam14a} can be extended to multiple recovery sets. We observe that
codes on Hermitian curves give a natural construction of LRC(2) codes. Motivated by it, we present a general construction of codes with multiple recovery sets on curves and construct several general families of LRC(2) codes with small locality. We also show
that LRC$(t)$ codes can be constructed using automorphism groups of curves and give such an interpretation for some earlier examples in \cite{tam14a}.

Concluding the introduction, we point out another line of thought associated with RS codes. Confining ourselves to the cyclic case, we can phrase the study of code parameters in terms of the zeros of the code. In classical coding theory 
this point of view leads to a number of nontrivial results for {\em subfield subcodes} of RS codes such as BCH codes and related code families. A similar study can be performed for LRC RS codes, with the main outcome being a characterization of both the distance and locality in terms of the zeros of the code. This point of view is further developed in \cite{tam15a}.

A part of the results of this paper were presented at the 2015 IEEE International Symposium on Information Theory and published in
\cite{barg15a}. The new results obtained in this paper include the extension to codes correcting locally more than one erasure (Theorem \ref{thm:rho} and related results) and the results on multiple recovery sets (Sect.~\ref{sect:avail} and related asymptotic
results).

\section{LRC Reed-Solomon Codes}

To prepare ground for the construction of LRC codes on curves let us briefly recall the construction of \cite{tam14a}.
Our aim is to construct an LRC code over $\ff_q$ with the parameters $(n,k,r)$, 
where $n\le q.$ We additionally assume that $(r+1)|n$ and $r|k$, although both the constraints can be lifted
by making adjustments to the construction described in \cite{tam14a}. 
\textcolor{black}{Let $A=\{P_1,\dots,P_n\}\subset \ff_q$ be a subset of points of $\ff_q$ and let $g(x)\in \ff_q[x]$ be
a polynomial of degree $r+1$ such that there exists a partition $\cA=\{A_1,\dots,A_{\frac n{r+1}}\}$
of  $A$ into subsets of size $r+1$ with the property that $g$ is constant on each 
of the sets $A_i\in \cA.$}

Consider the $k$-dimensional linear subspace $V\subset \ff_q[x]$ generated by the set of polynomials
   \begin{equation}\label{eq:basis}
     (g(x)^j x^i,\; i=0,\dots, r-1; j=0,\dots,\frac kr-1).
   \end{equation}
Given 
  \begin{equation}\label{eq:message}
a=(a_{ij},i=0,\dots,r-1;j=0,\dots,\frac kr-1)\in \ff_q^k
  \end{equation}
    let 
   \begin{equation}
   f_a(x)=\sum_{i=0}^{r-1}\sum_{j=0}^{\frac kr-1} a_{ij}x^ig(x)^j. 
     \label{eq:fa}
   \end{equation}
Now define the code $\cC$ as the image of the linear evaluation map
  \begin{equation}\label{eq:cc}
  \begin{aligned}
    e:V&\to\ff_q^n\\
    f_a&\mapsto (f_a(P_i), i=1,\dots,n).  
  \end{aligned}
  \end{equation}
As shown in \cite{tam14a}, $\cC$ is an $(n,k,r)$ LRC code whose minimum distance $d$ meets the bound \eqref{eq:dist}
with equality.  In particular, the locality property of the code $\cC$ is justified as follows.
Suppose that the erased coordinate $P$ is located in the set $A_i\subset [n].$ Note that the restriction $\delta_i(x)$
of the polynomial $f_a(x)$ to the set $A_i$ is a polynomial of degree at most $r-1$. Therefore, $\delta_i(x)$ 
can be found
by polynomial interpolation through the remaining $r$ coordinates of the set $A_i$. Once $\delta_i(x)$ is computed,
we find the value of the erased coordinate as $\delta_i(P).$

This construction can be also modified to yield $(n,k,r,\rho)$ LRC codes with arbitrary local distance defined above. Assume that
$(r+\rho-1)|n$ and $r|k$ and let $m=n/(r+\rho-1)$. Let $\cA=\{A_1,\dots,A_m\}$ be a partition of the set $A$ into
subsets of size $r+\rho-1$. Let $g\in \ff_q[x]$ be a polynomial of degree $r+\rho-1$ that is constant on
each of the sets $A_i$. We again represent the message vector $a$ in the form $\eqref{eq:basis}$ and map it to
the codeword using \eqref{eq:fa}-\eqref{eq:cc}. As shown in \cite{tam14a}, the obtained code has the parameters that meet the bound
\eqref{eq:rho} with equality. As above, the restriction $\delta_i(x)$ of the polynomial $f_a(x)$ to the subset $A_i$ has degree
at most $r-1,$ so it can found from any $r$ out of the $r+\rho-1$ coordinates in $A_i.$ Once $\delta_i$
is computed, it gives the values of all the remaining $\rho-1$ coordinates in $A_i.$

To construct examples of codes using this approach we need to find polynomials and partitions of points of 
the field that satisfy the above assumptions. As shown in \cite{tam14a}, one can take $g(x)=\prod_{\beta\in H}(x-\beta),$ where $H$ is a subgroup of the additive or the multiplicative group of $\ff_q$ (see also the 
example in the next section).
In this case $r=|H|-\rho+1,$ and the corresponding set of points $A$ can be taken to be any collection of the cosets
of the subgroup $H$ in the full group of points. In this way we can construct codes of length
$n=m(r+\rho-1),$ where $\rho\ge 2$ and $m\ge 1$ is an integer that does not exceed $(q-1)/|H|=(q-1)/(r+\rho-1)$ or $q/|H|{=q/(r+\rho-1)}$ depending on the choice
of the group. 

Codes in the family \eqref{eq:cc} in some cases also support the availability property. For instance to construct LRC(2) codes
one can take two subgroups $H_1,H_2$ in the group of points with trivial intersection. Let $|H_i|=r_i+1, i=1,2.$
\textcolor{black}{To construct the code using the above approach, we proceed as follows. Consider the polynomial algebras $\cP_1,\cP_2$ formed by the polynomials constant on the cosets of $H_i,i=1,2,$ 
respectively, and form the linear space $\cF=\oplus_{i=0}^{r_1-1}\cP_1x^i\cap \oplus_{j=0}^{r_2-1}\cP_1x^j.$ 
In this case  for any subspace $V\in \cF$  the evaluation code given by \eqref{eq:cc} has two disjoint recovery sets of 
size $r_i,i=1,2$ for each coordinate $i\in[n].$}

\section{Algebraic geometric LRC codes}\label{sect:AG}
In this section we present a general construction of LRC codes on algebraic curves.
As above, let us fix a finite field $\kk ={\ff}_q, q=p^a$ of characteristic $p$. To motivate our construction,
consider the following example.

{\em Example 1:} Let $H$ be a cyclic subgroup of $\ff_{13}^\ast$ generated by $3$ and let $g(x)=x^3.$ Let $r=2,n=9,k=4,$ and choose
$A=\{1,2,3,4,5,6,9,10,12\}.$ We obtain
  $\cA=\{A_1,A_2.A_3\},$ where
  \begin{equation}\label{eq:PP}
  \begin{aligned}
  A_1=\{1,3,9\},\;&&A_2=\{2,6,5\},\;&&A_3=\{4,12,10\}\\
      g(A_1)=1 & &g(A_2)=8 &&g(A_3)=12
  \end{aligned}
  \end{equation}
Note that the set $A_1$ forms the group of cube roots of unity in $\ff_{13}$ and that $A_2$ and $A_3$ are two of its cosets in 
$\ff_{13}^\ast.$

The set of polynomials \eqref{eq:basis} has the form $(1,x,x^3,x^4).$
In this case Construction \eqref{eq:cc} yields a $(9,4,2)$ LRC code with distance $d=5$ \cite{tam14a}.

This construction can be given the following geometric interpretation: the polynomial $g$ defines a covering map $g:{\mathbb P}^1 \to {\mathbb P}^1$ of degree $r+1=3$ such that the preimage of every point in $g(A)$ consists of
``rational'' points (i.e., $\ff_q$-points). 
This suggests a generalization of the construction to algebraic curves which we proceed
to describe (note Example 2 below that may make it easier to understand the general case).

Let $X$ and $Y$ be smooth projective absolutely irreducible curves over $\kk$. Let $g:X\to Y$ be a rational
separable map of curves of degree $r+1.$ As usual, denote by $\kk(X)$ ($\kk(Y)$) the field of rational functions on $X$ (resp., $Y$).
Let $g^\ast:\kk(Y)\to \kk(X)$ be the function that acts on $\kk(Y)$ by
$g^\ast(f)(P)=f(g(P)),$ where $f\in \kk(Y), P\in X.$  The map $g^\ast$ defines a field embedding $\kk(Y)\hookrightarrow \kk(X),$ and we identify $\kk(Y)$ with its image $g^*( \kk(Y))\subset \kk(X).$  

Since $g$ is separable, the primitive element
theorem implies that there exists a function $x\in \kk(X)$ such that $\kk(X)=\kk(Y)(x)$, and that satisfies
the equation 
  \begin{equation}\label{eq:x}
  x^{r+1}+b_r x^r+\dots+b_0=0,
  \end{equation}
  where $b_i\in \kk(Y).$
The function $x$ can be considered as a map $x: X\rightarrow \mathbb{P}^1_\kk,$ and we denote its degree $\deg (x)$ by $h.$ 

{\em Example 1:} (continued) For instance, in the above example, we have $X={\mathbb P}^1, Y={\mathbb P}^1,$ and the mapping $g$ is given by $y=x^3.$ We obtain $\kk(Y)=\kk(x^3)=\kk(y),$ $\kk(X)=\kk(y)(x),$ where $x$ satisfies the equation 
$x^3-y=0.$ Note that in this case $b_r=b_{r-1}=...=b_1=0,b_0=-y.$

The codes that we construct belong to the class of evaluation codes. Let 
$S=\{P_1,\dots,P_s\}\subset Y(\kk)$ be a subset of $\ff_q$-rational points of $Y$ and let $D$ be a 
positive divisor of degree $\ell\ge 1$ whose support is disjoint from $S.$ For instance, one can assume that 
$D\subset\pi^{-1}(\infty)$ for
a projection $\pi:Y\to \mathbb{P}^1_\kk.$ To construct our 
codes we introduce the following set of fundamental assumptions with respect to $S$ and $g$:
    \begin{gather}\label{eq:partition}
 A:= g^{-1}(S)=\{P_{ij}, i=0,\dots, r, j=1,\dots,s\}\subseteq X(\kk); \\
    g(P_{ij})=P_j \text{ for all } i,j; \notag \\
    b_i\in  L(n_iD), \quad i=0,1,\dots,r, \notag
    \end{gather}
for some natural numbers $n_i.$ 

Now let $\{f_1,\ldots, f_m\}$ be a basis of the linear space $L(D).$ 
The functions $f_i, i=1,\dots,m$ are contained in $\kk(Y)$ and therefore are constant on the fibers of the map $g$.
The Riemann-Roch theorem implies that $m\ge  \ell-g_Y+1,$ where $g_Y$ is the genus of $Y.$
  Consider the $\kk$-subspace $V$ of $\kk(X)$ of dimension $rm$ generated by the functions
  \begin{equation}\label{eq:fs}
 \{f_jx^i, i=0,\ldots,r-1,j=1,\ldots,m\}
  \end{equation}
  (note an analogy with \eqref{eq:basis}). Since $D$ is disjoint from $S$, the evaluation  map 
  \begin{equation}\label{eq:eval}
  \begin{aligned}
   e:=ev_A: &V\longrightarrow \kk^{(r+1)s}\\ 
     &F\mapsto (F(P_{ij}),i=0,\dots, r, j=1,\dots,s)
   \end{aligned}
  \end{equation}
 is well defined. The image of this mapping is a linear subspace of $\ff_q^{(r+1)s}$ (i.e., a code), which we denote 
by $\cC(D,g).$ The code coordinates are naturally partitioned into $s$ subsets $A_j=\{P_{ij},i=0,\dots,r\}, j=1,\dots,s$ of size $r+1$ each; see \eqref{eq:partition}. Assume throughout that, for any fixed $j$,  $x$ takes different values at the points in the set $(P_{ij},i=0,\dots,r).$
\begin{theorem}\label{thm:LRC} The subspace $\cC(D,g)\subset \ff_q$ forms an $(n,k,r)$ linear LRC code with the parameters
   \begin{equation}
\left.\begin{array}{c}
 n=(r+1)s  \\[.05in] k=rm\ge r(\ell-g_Y+1) \\ [.05in]
 d\ge n-\ell(r+1)-(r-1)h\end{array}\right\} \label{eq:d}
 \end{equation}
provided that the right-hand side of the inequality for $d$ is a positive integer.  Local recovery of an erased symbol $c_{ij}=F(P_{ij})$
can be performed by polynomial interpolation through the points of the recovery set $A_j$.
\end{theorem}
\begin{IEEEproof}
The first relation in \eqref{eq:d} follows by construction. The inequality for the distance is also immediate:
the function $f_j x^i,f_j\in L(D)$, evaluated on $A,$  can have at most $\ell(r+1)+(r-1)\deg (x)$ zeros. 
Since we assume that $d\ge1,$ the mapping $ev_A$ is injective, which implies the claim about the dimension of the code.
Finally, the functions $f_i$ are constant on the fibers $(P_{ij},i=0,\dots,r-1);$ therefore on each subset $A_j$ the
codeword is obtained as an evaluation of the polynomial of the variable $x$ of degree $\le r-1.$ This representation accounts
for the fact that coordinate $c_P, P\in A_j$ of the codeword can be found by interpolating a polynomial of degree at most $r-1$ through the remaining points of $A_j.$
\end{IEEEproof} 

The construction presented above can be extended to yield $(n,k,r,\rho)$ LRC codes, where $\rho\ge 3.$ Indeed, 
starting again with the curves $X$ and $Y,$ let us take $g:X\to Y$ to be a rational separable map of degree $r+\rho-1.$
Then the function $x\in \kk(X)$ such that $\kk(X)=\kk(Y)(x)$ satisfies an equation of degree $r+\rho-1$ (cf.~\eqref{eq:x}).
We again denote the degree of $x$ by $h$ and assume that $x$ is injective on the fibers. Let $S=\{P_1,\dots P_s\}\subset Y(\kk)$ and suppose that
$g$ is constant on the fibers of size $r+\rho-1$ lying above each of the points in $S,$ which form the
recovery sets $A_j=\{P_{ij}, i=0,\dots, r+\rho-2\}.$

Following the steps of the construction above and making obvious adjustments, we obtain a code $\cC_\rho(D,g)$ defined by
the evaluation map
 \begin{equation}\label{eq:eval-rho}
  \begin{aligned}
   e:=ev_A: &V\longrightarrow \kk^{(r+\rho-1)s}\\ 
     &F\mapsto (F(P_{ij}),i=0,\dots, r+\rho-2, j=1,\dots,s).
   \end{aligned}
  \end{equation}
\begin{theorem}\label{thm:rho}  The code $\cC_\rho(D,g)$ is an $(n,k,r,\rho)$ linear LRC code with the parameters
  \begin{equation}
\left.\begin{array}{c}
 n=(r+\rho-1)s  \\[.05in] k\ge r(\ell-g_Y+1) \\ [.05in]
 d\ge n-\ell(r+\rho-1)-(r-1)h\end{array}\right\} \label{eq:d-rho}
    \end{equation}
provided that the right-hand side of the inequality for $d$ is a positive integer. Local recovery of any $\rho-1$ symbols
that are contained in a single recovery set $A_j$, can be performed by polynomial interpolation through the remaining
$r$ points of this set.
\end{theorem}

 \vspace*{-.1in}\section{Some code families} 
 Let us give some examples of code families arising from our construction.
 
 \subsection{LRC codes from Hermitian curves}\label{sect:Herm}
 Let $q=q_0^2,$ where $q_0$ is a power of a prime, let $\kk=\ff_q,$ and
let $X:=H$ be the Hermitian curve, i.e., a plane smooth curve of genus $g_0=q_0(q_0-1)/2$ with the affine equation
     $$
     X: x^{q_0}+x=y^{q_0+1}.
     $$
     The curve $X$ has $q_0^3+1=q\sqrt q+1$ rational points of which one is the infinite point and the remaining
$q_0^3$ are located in the affine plane. There are two slightly different ways of constructing Hermitian LRC codes.
 
 \subsubsection{\underline{Projection on $y$}}
Here we construct $q$-ary $(n,k,r=q_0-1)$ LRC codes.
Take $Y={\mathbb P}^1(\kk)$ and take $g$ to be  the natural projection defined by $g(x,y):=y,$
then the degree of $g$ is $q_0=r+1$ and the degree of $x$ is $h=q_0+1.$ We can write
$X(\kk)=g^{-1}(\kk)\bigcup Q_{\infty}'$ where $Q_{\infty}'\in X$ is the  unique point over $ {\infty}\in Y.$

Turning to the code construction, take $ S=\kk\subset \mathbb{P}^1$ 
and $D=\ell Q'_\infty$ for some $\ell\ge 1.$ We have
   $$
   L(D)=\Big\{\sum_{i=0}^\ell a_i y^i\Big\} \subset \kk[y].
   $$
 Following the general construction of the previous section, we obtain the following result.
 \begin{proposition} \label{prop:Herm}The construction of Theorem \ref{thm:LRC} gives a family of $q$-ary Hermitian LRC codes
 with the parameters
   \begin{gather}
   n=q_0^3, k=(\ell+1)(q_0-1), r=q_0-1 \nonumber \\
  d\ge n-\ell q_0-(q_0-2)(q_0+1).\label{eq:dh}
  \end{gather}
  \end{proposition}
{\em Example 2:} Let $q_0=3,q=9,\kk=\ff_9$ and consider the Hermitian curve $X$ of genus 3 given by the equation
    $
    x^3+x=y^4.
    $
The curve $X$ has 27 points in the finite plane, shown in Fig.1 below (here $\alpha^2=\alpha+1$ in $\ff_9$), and one point at infinity.
 \remove{ \begin{align}
   \;&(0,0),(\alpha^2,0),(\alpha^6,0); \notag\\
  &(\alpha,\beta),(\alpha^3,\beta),(\alpha^4,\beta), \;\beta=1,\alpha^2,\alpha^4,\alpha^6; \label{eq:points}\\
  &(1,\beta),(\alpha^5,\beta),(\alpha^7,\beta),\;\beta=\alpha,\alpha^3,\alpha^5,\alpha^7,\notag
  \end{align}
where $\alpha^2=\alpha+1$ in $\ff_9,$ and one point at infinity. }

   \nc{{\bl}}{\bullet}
 \begin{figure*}[ht]\hrule  
\hspace*{.5in}\begin{minipage}{.4\linewidth}
 \begin{gather*}
  \begin{array}{c@{\hspace*{.05in}}c@{\hspace*{.05in}}*{10}{c@{\hspace*{.05in}}}}
   &\alpha^7&    &    &\bullet  &          &\bullet   &        &\bullet  &        &\bullet&\\  
   &\alpha^6&\bl &    &         &          &          &        &         &        &       &\\
   &\alpha^5&    &    &\bl      &          &\bl       &        &\bl      &        &\bl    &\\
   &\alpha^4&    &\bl &         &\bl       &          &\bl     &         &\bl     &       &\\
 x &\alpha^3&    &\bl &         &\bl       &          &\bl     &         &\bl     &       &\\
   &\alpha^2&\bl &    &         &          &          &        &         &        &       &\\
   &\alpha&      &\bl &         &\bl       &          &\bl     &         &\bl     &       &\\
   &1       &    &    &\bl      &          &\bl       &        &\bl      &        &\bl    &\\
   &0       &\bl &    &         &          &          &        &         &        &       &\\
     &&0 &1 &\alpha &\alpha^2 &\alpha^3 &\alpha^4 &\alpha^5 &\alpha^6 &\alpha^7&\\
     &&  &  &       &         &         &y
  \end{array}\\[.1in]
\text{\footnotesize Fig.1: 27 points of the Hermitian curve over $\ff_{9}.$ }
 \end{gather*}
\end{minipage}
\begin{minipage}{.4\linewidth}
 \begin{gather*}
  \begin{array}{c@{\hspace*{.05in}}c@{\hspace*{.1in}}*{10}{c@{\hspace*{.05in}}}}
   &\alpha^7&        &    &\alpha   &          &\alpha^7  &        &\alpha^5 &         &0     &\\  
   &\alpha^6&\alpha^2&    &         &          &          &        &         &         &      &\\
   &\alpha^5&        &    &\alpha^6 &          &\alpha^4  &        &\alpha^2 &         &0     &\\
   &\alpha^4&        &\alpha^7&     &\alpha^3  &          &\alpha^5&         &\alpha^5 &      &\\
x  &\alpha^3&        &\alpha^3&     &\alpha^7  &          &\alpha  &         &\alpha   &      &\\
   &\alpha^2&\alpha^3&    &         &          &          &        &         &         &      &\\
   &\alpha  &        &0   &         &0         &          &0       &         &0        &      &\\
   &1       &        &    &1        &          &\alpha^6  &        &\alpha^4 &         &0     &\\
   &0       &1       &    &         &          &          &        &         &         &      &\\[.05in]
          &&0       &1   &\alpha   &\alpha^2  &\alpha^3 &\alpha^4 &\alpha^5 &\alpha^6 &\alpha^7&\\
            &&  &  &       &         &         &y\vspace*{-.05in}
  \end{array}\\[.1in]
\text{\footnotesize Fig.2: Encoding of the message $(1,\alpha,\alpha^2,\alpha^3,\alpha^4,\alpha^5)$.}
  \end{gather*}
  \end{minipage}
  \vspace*{.05in}
  \hrule
\end{figure*}

\vspace*{.2in} \nd

The columns of the array in Fig. 1 correspond to the fibers of the mapping $g(\cdot,y)=y,$ 
and for every $a\in Y(\ff_9)\backslash Q_\infty$ there are 3 points $(\cdot,a)\in X$ lying above it.
These triples form the recovery sets $A_1,\dots,A_9,$ similarly to \eqref{eq:PP}.
  The map $x:X\to{\mathbb P}^1$ has degree $h=4.$ Choosing $D$ in the form $D=\ell Q'_\infty$ and taking $S=\ff_9$ 
(all the affine points of $Y$), we obtain an LRC code $\cC(D,g)$ with the parameters 
   \begin{gather}
    n=27, k=2(\ell+1), r=2\\
     d\ge 27-3\ell-4=23-3\ell, \quad \ell\ge 1.
     \end{gather}
 
For instance, take $\ell=2.$ The basis of functions \eqref{eq:fs} in this case takes the following form:
  \begin{equation*}
  \{1,y,y^2,x,xy,xy^2\}.
  \end{equation*}
To give an example of local decoding, let us compute the codeword for the message vector $(1,\alpha,\alpha^2,\alpha^3,\alpha^4,\alpha^5).$ 
The polynomial
$$   
  F (x,y)=1+\alpha y+\alpha^2y^2+\alpha^3x+\alpha^4xy+\alpha^5xy^2
  $$
evaluates to the codeword shown in Fig.~2 (e.g., $F(0,0)=1$, etc.).
Suppose that the value at $P=(\alpha,1)$ is erased. The recovery set for the coordinate $P$ is $\{(\alpha^4,1),(\alpha^3,1)\},$
so we compute a linear polynomial $f(x)$ such that $f(\alpha^4)=\alpha^7$ and $f(\alpha^3)=\alpha^3,$ i.e., 
$f(x)=\alpha x-\alpha^2.$ Now the coordinate at $(\alpha,1)$ can be found as $f(\alpha)=0.$\hfill\qed
 
  Computing the gap to the Singleton bound \eqref{eq:dist}, we obtain
   \begin{align}
   d+\frac kr(r+1)&\ge q_0^3-\ell q_0-(q_0-2)(q_0+1)+q_0(\ell+1)\nonumber\\
       &=q_0^3-q_0^2+2q_0+2\nonumber\\
       &=n-q+2\sqrt q +2.\label{eq:sg}
       \end{align}
 For codes that meet the bound \eqref{eq:dist} we would have $d+k(r+1)/r=n+2,$
 so the Singleton gap of the Hermitian LRC codes is {at most} $q-2\sqrt q=q_0(q_0-2).$ Of course, these codes cannot be Singleton-optimal because their length 
 is much greater than the alphabet size, but the gap in this case is still rather small.
 For instance in Example 2 we have $d+k(r+1)/r\ge 23-3\ell+3(\ell+1)=26$ while for codes meeting the Singleton bound we would have 
 $d+k(r+1)/r=29.$ 

As observed in \cite{ballico15}, the distance estimate of Prop.~\ref{prop:Herm} for some $\ell$ can be improved. Specifically,
if $q-q_0+1\le \ell\le q-1,$ the distance is bounded below as follows:
   $$
   d\ge q-\ell+1,
   $$
which is better than the estimate \eqref{eq:dh} if $q-q_0+1\le \ell\le q-1.$

\subsubsection{\underline{Projection on $x$}}\label{sect:Herm2} Again take $Y=\pp^1$ and let $g'(x,y):=x$ be the second natural projection on $\pp^1.$
There are $q_0$ points on $\pp^1$ that are fully ramified (they have only one point of $X$ above them), namely the points in the set
  \begin{equation}\label{eq:M}
  M=\{a\in \ff_q: a^{q_0}+a=0\}
  \end{equation}
(e.g., in Fig.~1 $M=\{0,\alpha^2,\alpha^6\}$). Therefore, every fiber of $g'$ over $\ff_q\backslash M$ consists of $\ff_q$-rational
points since there are in total
   $$
   |\ff_q\backslash M|\cdot(q_0+1)=q_0^3-q_0
   $$
rational points in those fibers.    Obviously $|g^{-1}(a)\cap (g')^{-1}(b)|\le1$ for all $a,b\in \ff_q.$

Take $S=\ff_q\backslash M$, then $r=q_0,$ and clearly $h=\deg(y)=q_0.$
We obtain
  \begin{proposition}
  The construction of Theorem \ref{thm:LRC} gives a family of $q$-ary Hermitian LRC codes
 with the parameters
   \begin{gather*}
   n=q_0^3-q_0, k=(\ell+1)q_0, r=q_0\\
  d\ge n-\ell(q_0+1)-q_0(q_0-1), \quad \ell\ge 1.
  \end{gather*}
  \end{proposition}
  For instance, in Example 2, taking $\ell=2,$ we obtain a code of dimension 9 from the basis of functions
  $
    \{1,y,y^2,x,xy,xy^2,x^2,x^2y,x^2y^2\}.
    $

Performing a calculation similar to \eqref{eq:sg} we obtain the quantity one less than for the first family:
   $$
   d+\frac kr(r+1)=n-q+2\sqrt q+1.
   $$
   
\begin{remark} Hermitian LRC codes are in a certain sense optimal for our construction. Note that 
most known curves with the optimal quotient (number of rational points)/(genus) have the property that
for any projection $g:X\to \pp^1$ the point $\infty\in \pp^1$ is totally
ramified (see e.g., the next section). In this case the quantity $h$ satisfies $h\ge n/q.$ At the same time, for
Hermitian curves, $h=n/q$ (or $(n/q)+1$). Recall also that Hermitian curves are absolutely maximal,
i.e. attain the equality in Weil's inequality, and moreover, their genus is maximal for
maximal curves.
\end{remark}

\subsection{LRC codes on Garcia-Stichtenoth curves}\label{sect:GScodes}
 
  Let $q=q_0^2$ be a square and let $l\ge 2$ be an integer. Define the curve $X_l$ and the functions $x_l,z_l$ inductively as follows:
    \begin{gather}
 x_0:=1;\;X_1:= \mathbb{P}^1, \kk(X_1)=\kk(x_1); \label{eq:GST1}\\
X_l: z_l^{q_0}+z_l=x_{l-1}^{q_0+1}, \text{ where for $l\ge 3$, }x_{l-1}:=\frac{z_{l-1}}{x_{l-2}} \in \kk(X_{l-1}) ,\notag
  \end{gather}
  where $\kk=\ff_q.$
In particular, $X_2=H$ is the Hermitian curve.
 The resulting family of curves is known to be asymptotically maximal \cite{garcia95}, \cite[p.177]{TVN07}, 
 and gives rise to codes with good parameters in the standard error correction problem. 
Since this family generalizes Hermitian curves, we can expect that it gives rise to two families of codes that extend
the constructions of Sect.~\ref{sect:Herm}. This is indeed the case, as shown below.

\subsubsection{}\label{sect:GS1}
To use the general construction that leads to Theorem \ref{thm:LRC} we take the map $g_l:X_l\to X_{l-1}$ to be the natural projection  of
degree $q_0=r+1.$ We note that
    \begin{equation}\label{eq:gs1}
        g_l^\ast: \kk(X_{l-1})\to\kk(X_l)=\kk(X_{l-1})(z_l).
    \end{equation}
To describe rational points of the curve $X_l$ let $\psi_l:X_l\to {\mathbb P}^1$ be the natural projection of degree $q_0^{l-1},$
i.e., the map $\psi_l=g_l\circ g_{l-1}\circ\dots\circ g_2.$ Then all the points in the preimage $\cP_l:=\psi_l^{-1}(\ff_q^\ast)$ are
$\ff_q$-rational, and there are $n_l=q_0^{l-1}(q_0^2-1)$ such points.
The genus of the curve $X_l$ can be bounded above as
     $$
     G_l\le q_0^{l}+q_0^{l-1}=q_0^{l-1}(q_0+1)=\frac{n_l}{q_0-1}
     $$
(the exact value of $G_l$ is known \cite{garcia95}, but this estimate suffices: in particular, it implies that the curves $X_l, l\to \infty$
are asymptotically maximal). 
We obtain the following result.
 \begin{proposition}\label{prop:GS1}  There exists 
 a family of $q$-ary $(n,k,r=q_0-1)$ LRC codes on the curve $X_l, l\ge 2$  with the parameters
   \begin{equation}\label{eq:GS1}
   \left.
   \begin{array}{c}
   n=n_l=q_0^{l-1}(q_0^2-1)\\[.1in]
   \displaystyle k\ge r\Big(\ell-\frac{n_{l-1}}{q_0-1}+1\Big) \\[.1in]
  \displaystyle d\ge n_l-\ell q_0-\frac{2n_l(q_0-2)}{q_0^2-1}
  \end{array}\right\}
  \end{equation}
  where $\ell$ is any integer such that $G_{l-1}\le \ell\le n_{l-1}.$
  \end{proposition}
  \begin{IEEEproof}
We apply the construction of Theorem \ref{thm:LRC} to $X:=X_l, Y:=X_{l-1},$ taking the map $g:=g_l, $ $Q_\infty:=P_{\infty,l},$ $D=\ell Q_{\infty}.$

The function $x$ in the general construction in this case is $x=z_l.$ To estimate the distance of the code $\cC(D,g)$ using \eqref{eq:d} we
need to find the degree $h=\deg(z_l).$ Toward this end, observe that $z_l=x_lx_{l-1},$ so 
   $$
    \deg(z_l)=\deg(x_l)+\deg(x_{l-1}).
    $$
Let $(x_l)_0^{(l)}$ be the divisor of zeros of $x_l$ on $X_l.$ Recall from \cite{garcia95}, Lemma 2.9 that  $(x_l)_0^{(l)}=q_0^{l-1}Q_l,$
where $Q_l$ is the unique common zero of $x_1,z_2,\dots,z_l.$ Therefore, $\deg(x_l)_0^{(l)}=q_0^{l-1}$ and $\deg(x_{l-1})_0^{(l-1)}=q_0^{l-2}.$
Since the map $X_l\to X_{l-1}$ is of degree $q_0,$ we obtain $\deg (x_{l-1})_0^{(l)}=q_0\deg(x_{l-1})_0^{(l-1)}=q_0^{l-1}.$
Summarizing, 
 $$
   h=2q_0^{l-1}=\frac{2n_{l}}{q_0^2-1}.
 $$
Now the parameters in \eqref{eq:GS1} are obtained from \eqref{eq:d} by direct computation. 
\end{IEEEproof}
 
\subsubsection{}\label{sect:GS2} Now consider the second natural projection of curves in the tower \eqref{eq:GST1}. Namely, let
$Y_l$ correspond to the function field $\kk(z_2,\dots,z_l)$ and consider the field embedding
   $$
      (g_l')^\ast:\kk(Y_l)\to \kk(X_l)=\kk(x_1,z_2,\dots,z_l).
      $$
Note that $g_2'$ is the projection $g':X_2\to\pp^1$ considered in Section \ref{sect:Herm2}. The curves
$\{Y_l,l=2,3,\dots\}$ form another optimal tower of curves \cite[Remark 3.11]{garcia96} given by the recursive equations
   $$ 
   Y_{l}: z_{l}^q +z_{l} =\frac{z_{l-1}^q }{z_{l-1}^{q-1} + 1}, \;l\ge 3;\; 
Y_2:=\mathbb{P}^1.
   $$
In geometric terms, the embedding $(g'_l)^\ast$ implies that 
the curve $X_l$ is the fiber product of $X_2$ and $Y_l$ over $Y_2=\mathbb{P}^1,$ viz.
$X_l= X_2\times_{Y_2}Y_l,$ which in turn implies that the projection 
$g'_l: X_{l}\rightarrow  Y_l$ shares the main properties of $g'=g'_2$. Indeed, we have:
\begin{enumerate}
  \item The genus of $Y_l$ satisfies $G_l'<q_0^{l-1}$ (the exact value is given in \cite[Remark 3.8]{garcia96}; note that the
  notation for $\kk(Y_l)$ in \cite{garcia96} is $T_{l-1}).$
  \item Let $\pi_l:Y_l\to Y_2$ be the natural projection of degree $\deg(\pi_l)=q_0^{l-2}$. All the points in $S_l:=\pi_l^{-1}(\ff_q\backslash
  M)$ are $\ff_q$-rational and
     $$
  |S_l|=q_0^{l-2}(q_0^2-q_0)=q_0^{l-1}(q_0-1)=n_l/(q_0+1).
     $$
 \item The point $\infty\in Y_2=\pp^1$ is totally ramified, i.e., $\pi_l^{-1}(\infty)=P_{\infty,l}'$
     for a rational point $P_{\infty,l}'\in Y_l.$
 \item We have $(g'_l)^{-1}(S_l)= (\psi_l)^{-1}(\ff_q), |(g'_l)^{-1}(S_l)|=n_l,$ and all the points in $(g'_l)^{-1}(S_l)$ are $\ff_q$-rational.  The degree of the projection $g_l'$ is $ \deg (g'_l)=q_0+1.$ The fibers of $g'_l$ are transversal with those of  $g_l.$
\item The degree of $x_1: X_l\longrightarrow \mathbb{P}^1$ equals 
$h:=\deg (x_1)= \deg (\pi_l)\deg ((x_1)_0^{(2)})=q_0^{l-1}.$          
  \end{enumerate}
We obtain the following statement.
\begin{proposition}\label{prop:GS2}
There exists  a family of $q$-ary $(n,k,r=q_0)$ LRC codes on the curve $X_l, l\ge 2$  with the parameters
   \begin{equation}\label{eq:GS2}
   \left.
   \begin{array}{c}
   n=n_l=q_0^{l-1}(q_0^2-1)\\[.1in]
   \displaystyle k\ge r(\ell-q_0^{l-1}+1) \\[.1in]
  \displaystyle d\ge n_l-\ell (q_0+1)-(q_0-1)q_0^{l-1}
  \end{array}\right\}
  \end{equation}
  where $\ell$ is any integer such that $G_{l-1}\le \ell \le n_{l-1}.$
\end{proposition}  
{\em Proof:} Put $r=q_0$ and apply the construction of Theorem \ref{thm:LRC} to 
   $$
X:=X_l, Y:=Y_l, g:=g_l', Q_\infty:=P_{\infty,l}', D=\ell Q_{\infty}.\hspace*{.3in}\qed
   $$
   
{\em Remark 4.2:} For the construction of Prop.~\ref{prop:GS2} the lower bound of Remark 4.1
takes the form $h\ge n_l/q_0^2=q_0^{l-1}-q_0^{l-3}$ which is very close the actual value $h=q_0^{l-1}.$
In the case of Prop. \ref{prop:GS1} the value $h$ is about twice as large as the lower bound. 

{\em Remark 4.3:} Due to the results of \cite{shum01}, the basis of the function space $L(D_t)$ and the set $S_l$ can be
found in time polynomial in $n_l$, and so the codes of Prop.~\ref{prop:GS2} are polynomially constructible.
 
\subsection{Modifications of the main construction: Small locality}
The constructions of the previous section yield infinite families of $q$-ary LRC codes with good parameters. At the same time, they are
somewhat rigid in the sense  that the locality parameter $r$ fixed and is equal to about $\sqrt q.$ Generally one would prefer to construct LRC codes for any given $r$, or at least for a range of its values.
It is possible to modify the above construction to attain small locality (such as, for instance, $r=2$), while still
obtaining code families that improve upon the GV bound \eqref{eq:GV}. 

We again begin with the Garcia-Stichtenoth tower of curves $X_l$ given by \eqref{eq:GST1}. The codes that we construct
will have locality $r$, where $(r+1)|(q_0+1)$. Let $X:=X_l$ and let $Y:=Y_{l,r}$ be the curve with the function field
  $$
    \kk(Y_{l,r}):=\kk(x_1^{{r+1}},z_2,\dots,z_l).
  $$ 
Consider a covering map $g:X\to Y$ defined by the natural projection $x_1\mapsto x_1^{{r+1}}$.
Using the pair $(X,Y)$ in Theorem \ref{thm:LRC}, we obtain the following result.
\remove{
We obtain the code parameters   
  \begin{equation}\label{eq:GS3}
   \left.
   \begin{array}{c}
   n=n_l=q_0^{l-1}(q_0^2-1)\\[.1in]
   k\ge r(t-g_Y+1) \\[.1in]
   d\ge n_l-t (r+1)-(r-1)q_0^{l-1}
  \end{array}\right\}
  \end{equation}}
  
\begin{proposition}\label{prop:GS4} Let $(r+1)|(q_0+1)$. There exists  a family of $q$-ary $(n,k,r)$ LRC codes on the curve 
$X_l, l\ge 2$  with the parameters
   \begin{equation}\label{eq:GS4}
   \left.
   \begin{array}{c}
   n=n_l=q_0^{l-1}(q_0^2-1)\\[.1in]
   \displaystyle k\ge r\Big(\ell-q_0^{l-1}\frac{q_0+1}{r+1}+1\Big) \\[.1in]
  \displaystyle d\ge n_l-\ell  (r+1)-(r-1)q_0^{l-1}  \end{array}\right\}
  \end{equation}
  where $\ell $ is any integer such that $g_{Y}\le \ell \le n_{l-1}.$
\end{proposition}  
{\em Proof:}   Recall that the Riemann-Hurwitz formula \cite[p.102]{TVN07} implies that for any (surjective) covering 
$f: X\longrightarrow Y$ of degree $N$ between smooth absolutely irreducible curves one has the inequality $g_X\ge 1+N
(g_Y-1),$ or $g_Y\le 1+\frac{g_X-1}{N}.$ Applying this inequality to our pair of curves we get an upper estimate of  the genus of  $Y$, and the
 parameters of the code are obtained directly from \eqref{eq:d}.
\qed

\subsection{Modifications of the main construction: Correcting more than one erasure}
Another modification relates to $(n,k,r,\rho)$ LRC codes constructed in Theorem \ref{eq:eval-rho}, where $\rho\ge 2$. 
It is possible to adjust the code families constructed above in this section to address this case.
For instance, retracing the 
steps that lead to Propositions \ref{prop:GS1}, \ref{prop:GS2}, we can construct sequences of LRC codes that correct more than one erasure within a recovery set.

\vspace*{.1in}
\begin{proposition}\label{prop:GS2-rho} Let $q=q_0^2,$ where $q_0$ is a power of a prime. There exists  a family of $q$-ary $(n,k,r,\rho)$ LRC codes on the curve $X_l, l\ge 2$  
with $r+\rho-1=q_0$ whose parameters satisfy the following relations:
 \begin{equation}\label{eq:GSrho1}
   \left.
   \begin{array}{c}
   n=n_l=q_0^{l-1}(q_0^2-1)\\[.1in]
   \displaystyle k\ge r\Big(\ell-\frac{n_{l-1}}{q_0-1}+1\Big) \\[.1in]
  \displaystyle d\ge n_l-\ell q_0-\frac{2n_l(q_0-2)}{q_0^2-1}
  \end{array}\right\}
  \end{equation}
  There exists a family of $q$-ary $(n,k,r,\rho)$ LRC codes with $r+\rho-2=q_0$ and
  \begin{equation}\label{eq:GSrho2}
   \left.
   \begin{array}{c}
   n=n_l=q_0^{l-1}(q_0^2-1)\\[.1in]
   \displaystyle k\ge r(\ell-q_0^{l-1}+1) \\[.1in]
  \displaystyle d\ge n_l-\ell (q_0+1)-(q_0-1)q_0^{l-1}
  \end{array}\right\}
  \end{equation}
In both cases  $\ell$ is any integer such that $G_{l-1}\le \ell\le n_{l-1}.$
\end{proposition}
\vspace*{.1in}
Clearly, it is also possible to make a similar claim about the existence of $(n,k,r,\rho)$ codes relying on Proposition~\ref{prop:GS4}. 
We confine ourselves to these brief remarks, noting that similar results arise from codes on Hermitian curves as well as from the
other families mentioned in this paper.

\remove{   \begin{equation}\label{eq:GS2}
   \left.
   \begin{array}{c}
   n=n_l=q_0^{l-1}(q_0^2-1)\\[.1in]
   \displaystyle k\ge r(t-q_0^{l-1}+1) \\[.1in]
  \displaystyle d\ge n_l-t (q_0+1)-(q_0-1)q_0^{l-1}
  \end{array}\right\}
  \end{equation}
  where $t$ is any integer such that $G_{l-1}\le t\le n_{l-1}.$
}

\section{{The Availability Problem: Multiple recovery sets}} \label{sect:avail}
In this section we present a general construction of codes on curves with multiple recovery sets. To simplify the notation,
we restrict ourselves to the case $t=2$, but it will be seen that our approach extends immediately to any number $t$ of
recovery sets.

\subsection{An example}\label{sect:example} We begin with an example for Hermitian curves, which motivates the general description.
The existence of two projections $g$ and $g'$ with mutually transversal fibers suggests
that Hermitian LRC codes could be modified, leading to a family of LRC(2) codes
with two recovery sets of size $r_1=q_0-1$ and $r_2=q_0,$ respectively.
Indeed, let
   $$
   B=g^{-1}(\ff_q\backslash\{0\})=(g')^{-1}(\ff_q\backslash M)\subset X/\ff_q ,
   $$
$|B|=(q_0^2-1)q_0,$ where $M$ is defined in \eqref{eq:M}, and consider the following polynomial space of dimension $(q_0-1)q_0:$
   $$
   L:=\text{span\,}\{x^iy^j, i=0,1,\dots,r_1-1,j=0,1,\dots,r_2-1\}.
   $$
\begin{proposition} Consider the linear code $\cC$ obtained by evaluating the functions in $L$ at the points of $B$.
  The code $\cC$ has the parameters $(n=(q_0^2-1)q_0,k=(q_0-1)q_0,\{r_1=q_0-1,r_2=q_0\})$ and distance
    \begin{equation}\label{eq:distance}
      d\ge (q_0+1)(q_0^2-3q_0+3).
    \end{equation}
\end{proposition}
  \begin{IEEEproof} $X$ is a plane curve of degree $q_0+1$, so the Bezout theorem implies that any polynomial of degree $\le 2q_0-3$
  has no more than $(q_0+1)(2q_0-3)$ zeros on $X;$ hence \eqref{eq:distance}. All the other parts of the claim are obvious.\end{IEEEproof}
For instance, puncturing the code of Example 2 on the coordinates in $M,$ we obtain an LRC(2) code with the parameters
$(24,6,\{2,3\})$ and distance $d\ge 12.$  

\subsection{General construction}\label{sect:gen-availability}
The general construction of LRC(2) codes on curves can be described as follows. Let $X,Y,Y_1,$ and $Y_2$ be smooth projective
absolutely irreducible algebraic curves over $\kk$ defined together with regular surjective separable maps between them as shown in the following commutative diagram:

\begin{center}\begin{tikzcd}
&X \arrow{rd}[description]{g_2}\arrow{ld}[description]{g_1}\arrow{dd}[description]{g}\\
Y_1 \arrow{rd}[description]{h_1}&&Y_2 \arrow{ld}[description]{h_2}\\
&Y
\end{tikzcd}.
\end{center}

Here $g:X\to Y$ is a map degree $d_g$, the maps $g_1:X\to Y_1$ and $h_2:Y_2\to  Y$ are of degree $d_{1,g},$ and the maps
$g_2:X\to Y_2$ and $h_1:Y_1\to Y$ are of degree $d_{2,g}$. This implies that $d_g=d_{1,g}d_{2,g}$, and this construction
identifies $X$ with the fiber product of curves $Y_1\times_Y Y_2,$ which means that
   $$
   g^\ast(\kk(Y))=g_1^\ast (\kk(Y_1))\cap g_2^\ast(\kk(Y_2))
   $$
inside $\kk(X).$

We assume that the maps $g,g_1,g_2$ satisfy the following set of assumptions.

(i) Suppose that $\kk(X)=g_1^\ast(\kk(Y_1))(x_1), \kk(X)=g_2^\ast(\kk(Y_2))(x_2),$ and $\kk(X)=g^\ast(\kk(Y))(x_1,x_2),$
where $x_1,x_2$ are primitive elements of their respective separable extensions that satisfy conditions similar to those
discussed above (cf.~\eqref{eq:x}). As before, we can also write $\kk(X)=\kk(Y)(x),$ where $x$ is a primitive element
that generates $\kk(Y)$ over $\kk(X)$, and denote its degree by $h.$ 

(ii) Let $B\subset A\subset X(\kk)$, where $A$ is the set of the form defined in \eqref{eq:partition}. Assume that the subset $B$
can be partitioned into pairwise disjoint subsets in two different ways:
   $$
   B=\bigcup_{y_1\in g_1(B)} g_1^{-1}(y_1)=\bigcup_{y_2\in g_2(B)} g_2^{-1}(y_2)
$$
so that all the fibers of $g_1$ over $g_1(B)\subset Y_1(\kk)$ and of $g_2$ over $g_2(B)\subset Y_2(\kk)$ consist of $\kk$-rational
points of $X.$

(iii) Finally, assume that the fibers $g_1^{-1}(y_1)$ and $g_2^{-1}(y_2)$ for any $(y_1,y_2)\in g_1(B)\times g_2(B)$ are transversal,
i.e., $|g_1^{-1}(y_1)\cap g_2^{-1}(y_2)|\le 1.$

\vspace*{.1in}\begin{definition}{($\cC(D,g_1,g_2)$ code)} Let $D$ be a positive divisor on $Y$ of degree $\ell$ 
such that $\supp(D)\subset \pi^{-1}(\infty),$ and let $\{f_1,\dots,f_m\}$ be
a basis of the linear space $L(D)\subset \kk(Y)$. Consider the following polynomial space of dimension $md_g$:
   $$
   L:=\text{span\,}\{x_1^ix_2^jf_k, i=0,1,\dots,d_{1,g}-2,j=0,1,\dots,d_{2,g}-2,k=1,\dots,m\}\subset\kk(X).
   $$
The code $\cC(D,g_1,g_2)$ of length $n=|B|$ is constructed as the image of the evaluation map
   \begin{equation}\label{eq:eval1}
  \begin{aligned}
   e:=ev_B: &V\longrightarrow \kk^{|B|}\\ 
     &F\mapsto (F(b),b\in B)
   \end{aligned}
  \end{equation}
\end{definition}
Note that since $\supp(Q_\infty)$ is disjoint from $B$, this map is well defined. The properties of the code $\cC(D,g_1,g_2)$ are
summarized in the following theorem whose proof is analogous to Theorem \ref{thm:LRC}.

 \begin{theorem}\label{thm:LRC2} The subspace $\cC(D,g_1,g_2)\subset \ff_q^{|B|}$ forms an $(n,k,\{r_1:=d_{1,g}-1,r_2:=d_{2,g}-1\})$ linear LRC(2) code with the parameters
   \begin{equation}
\left.\begin{array}{c}
 n=|B|  \\[.05in] k=(d_g-1)m\ge (d_g-1)(\ell-g_Y+1) \\ [.05in]
 d\ge n-\ell d_g-(d_{1,g}+d_{2,g}-4)h\end{array}\right\} \label{eq:d1}
 \end{equation}
for any $\ell\ge 1,$ provided that the right-hand side of the inequality for $d$ is a positive integer.
\end{theorem}
We note that the choice of parameters in this construction is flexible because of many options for the choice of the degrees
of the maps $g_1$ and $g_2.$ Some examples illustrating this statement are given below.

\subsection{Group-theoretic construction}\label{sect:group}
Assumptions (i)-(iii) for the maps $g_1$ and $g_2$ can be satisfied in the following situation. Suppose that the automorphism group
${Aut}_{\kk}(X)$ of the curve $X$ contains a semi-direct product of two subgroups: 
$H_1,H_2\le G:=Aut_{\kk}(X)$ and $H=H_1\rtimes H_2.$  
In this case the curves $Y_1,Y_2,$ and $Y$ can be naturally defined by their function fields:
   $$ 
   \kk(Y_1):=  (\kk(X))^{H_1},\;\kk(Y_2):=  (\kk(X))^{H_2},\;\kk(Y):=  (\kk(X))^{H},\;
   $$
where $(\kk(X))^{K}$ for $K\le G$ is the subfield of elements invariant under $K$. 
Now suppose that the subset $B\subset X(\kk)$ in the general construction is a union of $H$-orbits. It is easy to check
that in this case the assumptions for the maps $g_1,g_1,$ and $g$ stated above are satisfied, and therefore we obtain
a general way of constructing LRC(2) codes.

In particular, the LRC(2) codes constructed in Examples 5, 6, 7 and Propositions 4.2, 4.3 of \cite{tam14a} are of this type for $X=\pp^1_{\kk}$ and appropriate subgroups $H_1$ and $H_2$ in  ${Aut}_{\kk}(\pp^1)={\mathrm PGL}_2(\kk)$ (in \cite{tam14a}
these subgroups are denoted by $H$ and $G$).

\begin{remark} [More than two recovery sets]
As noted above, the considerations of Sect.~\ref{sect:gen-availability}, \ref{sect:group} can be extended without difficulty to codes with
any number $t\ge 2$ of recovery sets. Nontrivial examples can be constructed for curves whose automorphism groups have many
subgroups, which is the case for many known families of good curves.
\end{remark}

\subsection{LRC(2) codes on Hermitian curves}\label{sect:gen-H} Let us use the code construction of the previous section to generalize the 
example of LRC(2) codes on Hermitian curves. 

 Let $X$ be the Hermitian curve over $\kk=\ff_q$. 
Let $e_1|(q_0+1)$ and consider the map $g_1:X\to Y_1$ of degree $d_1=\frac{q_0+1}{e_1}$ given by the projection $g_1(x,y):=(x,y^{d_1})$, and let $r_1=d_1-1$. The image of $g_1$ is the curve
  $$
   Y_1: x^{q_0}+x=u^{e_1}
     $$
with the function field $\kk(Y_1)=\kk(x,u:=y^{d_1}).$ Likewise, let $d_2|q_0$ be a divisor such that
$q_0=d_2^a$ for some natural number $a\ge 1$ and let $g_2:X\to Y_2$ be the projection
$g_2(x,y):=(v:=x^{d_2}+x,y)$ on the curve $Y_2$ with the function field $\kk(v,y).$ Let $r_2=d_2-1$ and $e_2=q_0/d_2=d_2^{a-1}.$
Define the curve $Y$ by $\kk(Y):=\kk(Y_1)\cap \kk(Y_2)\subset\kk(X).$
Using the notation in Sect.~\ref{sect:example}, let
   $$
   B:=g^{-1}(\ff_q\backslash\{0\})=(g_1)^{-1}(\ff_q\backslash M)
   $$
so $|B|=(q-1)q_0$, where the set $M$ is defined in \eqref{eq:M}.

\vspace*{.1in}
\begin{proposition} If $e_1=1$ or $e_2=1$ then $g_Y=0$. If $(q_0,e_1)\in\{(2,3),(3,2)\}$ then $g_{Y_1}=1$. If $q_0=2, e_2=2=d_2,$ then
$g_{Y_2}=1.$ Otherwise, 
the genera of the curves $Y,Y_1,Y_2$ are given by
  \begin{gather}
  g(Y_1)=\frac{(q_0-1)(e_1-1)}{2}, \quad
  g(Y_2)=\frac{q_0(d_2^{a-1}-1)}2,  \label{eq:g1}\\
  g(Y)\le \tilde g_Y:=\min\left\{\frac{e_2(e_1-1)}{2}-\frac{e_1+1}{2d_2}+1,\frac{e_1(e_2-1)}{2}-\frac{e_2+1}{2d_1}+1\right\}.
    \label{eq:gY}
  \end{gather}
\end{proposition}
\begin{IEEEproof}
The curves $Y_1,Y_2,$ and $Y$ are images of regular surjective maps of a maximal curve and therefore themselves maximal.
Let $Z$ be a maximal curve $Z$ over $\ff_{q_0^2}$ with $N_Z=|Z(\ff_{q_0^2})|$ rational points.
The Weil inequality gives
   $$
   g_Z=\frac{N_Z-q_0^2-1}{2q_0}
   $$
The maps $g_1$ and $g_2$ are subcovers of the $x$-projection and the $y$-projection, respectively (viz. Fig.~1). The 
description of the fibers of these projections given above implies that
   $$ N_{Y_1}=\frac{N_X-|M|}{d_1}+|M|=\frac{q_0^3-q_0}{d_1}+q_0+1=q_0(q_0-1)e_1+q_0+1,$$
   $$N_{Y_2}=\frac{N_X-1}{d_2}+1=\frac{q_0^3}{d_2}+1=q_0^2e_2+1,$$
and thus
      \begin{gather*} g({Y_1})=\frac{N_{Y_1}-q_0^2-1}{2q_0}=\frac{(q_0-1)(e_1-1)}{2}, 
      \\
      g({Y_2})=\frac{N_{Y_2}-q_0^2-1}{2q_0}=\frac{q_0(e_2-1)}{2}=\frac{q_0(d_2^{a-1}-1)}{2}.
      \end{gather*}
The cases of $e_i=1, i=1,2; e_2=2$ follow immediately. Finally, we use the Hurwitz formula \cite[p.102]{TVN07} to obtain
   $$
   g(Y)\le \min\Big\{\frac{g(Y_1)-1}{d_2}+1,\frac{g(Y_2)-1}{d_1}+1\Big\},
   $$
hence \eqref{eq:gY}.   
\end{IEEEproof}

To construct LRC(2) codes on Hermitian curves, we use this proposition together with Theorem \ref{thm:LRC2}. This yields 
the following code family.
\vspace*{.1in}
\begin{theorem} There exists a family of LRC(2) codes on Hermitian curves with the parameters
\begin{equation*}
\left.\begin{array}{c}
 n=|B|=(q-1)q_0  \\[.05in] k=r_1r_2m\ge r_1r_2 (\ell-\tilde g_Y+1) \\ [.05in]
 d\ge n-\ell d_1d_2-(d_{1}+d_{2}-4)q_0\end{array}\right\} 
 \end{equation*}
for $D:=\ell Q_{\infty}, $ where  $Q_{\infty}\in Y(\kk)$ is the unique point over $\infty\in P^1(\kk)$ and where
$\tilde g_Y$ is defined in \eqref{eq:gY}.
\end{theorem}

\vspace*{.1in}
Let us give a numerical example:
For $q_0=3^{2b+1},q=3^{4b+2}$ we can take  $d_1:=\frac{q_0+1}{e_1}=4=r_1+1,e_1=\frac{q_0+1}{4} ,\,d_2:=q_0/e_2=3=r_2+1, e_2=3^{2b}$ and $\; d_g=d_1d_2=12,$   

$$\tilde g_Y=\left \lfloor{ \frac{4\cdot 3^{4b+2}-13\cdot 3^{4b+1}+19}{24} }\right \rfloor.$$
For $b=1$ we get $n=3^3(3^6-1)=19656,k= 6m\ge 6(t-106),d\ge 19440-12t.$
In conclusion we note that this example can be also be treated in the framework of the group case considered in 
Sect.~\ref{sect:group}.

\subsection{LRC(2) codes on Garcia-Stichtenoth curves}
Let us use the approach developed in this section for the family of curves defined in \eqref{eq:GST1}. As before, let $q=q_0^2.$

As in Sect.~\ref{sect:gen-H}, let us take $e_1|(q_0+1), e_2|q_0$ and take the locality parameters $r_1,r_2$ given by
$\frac{q_0+1}{e_1}=r_1+1, \frac{q_0}{e_2}=r_2+1,$ where $q_0=d_2^a,e_2=d_2^{a-1} $ for some natural number $a$.
Define the curves $Y_{l,1},Y_{l,2}$ and $Y_l$ by 
  \begin{gather*}
   \kk(Y_{l,1}):=\kk(x_1^{{r_1+1}},z_2,\dots,z_l)\\
   \kk(Y_{l,2}):=\kk(X_{l-1},z_l^{d_2}+z)\\
   \kk(Y_l):=\kk(Y_{l,1})\cap \kk(Y_{l,2})\subset \kk(X_l)
  \end{gather*}
which naturally defines the corresponding projections $g_{1,l}:X_l\to Y_{l,1},$ $ g_{2,l} X_l\to Y_{l,1},$ and $g_l:X_l\to Y_{l}$ of 
respective degrees $d_1, d_2$ and $d_1d_2$. Note also that by the Hurwitz formula 
    \begin{equation}\label{eq:g}
    g(Y_{l})\le \frac{G_{l}-1}{d_1d_2}+1\le \frac{n_l}{(q_0-1)d_1d_2}+1= \frac{q_0^{l-1}(q_0+1)}{d_1d_2}+1= \frac{q_0^{l-1}(q_0+1)}{(r_1+1)(r_2+1)}+1.
    \end{equation}
Moreover since the family  $\{X_l,\, l=0,1,...\}$   is asymptotically optimal, so is the family of curves $\{Y_l,\, l=0,1,...\}.$
Therefore, by the Drinfeld-Vl\u adu\c t inequality \cite[p.~146]{TVN07} the genus $g(Y_l)$ asymptotically tends to the quantity on the right-hand side of \eqref{eq:g} for $l\to \infty.$ 

 The set $B$ in the construction \eqref{eq:eval1} can be chosen as 
   $$
   B:=\cP_l\subset X_l(\kk), \;|B|=q_0^{l-1}(q_0^2-1).
   $$ 
The general construction of Theorem \ref{thm:LRC2} gives the following result. 
  
\begin{proposition}\label{prop:GS2a} Let $D:=\ell Q_{\infty},$ then the code
$\cC(D,g_{1,l},g_{2,l})$ is a $q$-ary $(n_l,k,\{r_1,r_2\})$ LRC(2) code whose length, dimension, and distance satisfy the following
relations:
  \begin{equation*}
\left.\begin{array}{c}
 n_l=|B|=(q-1)q^{l-1}_0 =q_0^{l-1}(q_0^2-1) \\[.05in] 
 k=r_1r_2m\ge r_1r_2 \Big(\ell- \displaystyle{\frac{q_0^{l-1}(q_0+1)}{(r_1+1)(r_2+1)}}\Big) \\ [.05in]
 d\ge n-\ell d_1d_2-(d_{1}+d_{2}-4)q_0^{l-1}\, \end{array}\right\} 
 \end{equation*}
where $\ell\ge 1$ is any natural number such that the estimate for $d$ is nontrivial. 
\end{proposition}
  
\section{Asymptotic constructions}

In this section we consider asymptotic parameters of some code families constructed above. We derive asymptotic bounds
on the rate as a function of the relative distance for LRC codes that correct one or more erasures (see Def.~\ref{def:LRC}, 
\ref{def:LRC-rho}) as well as for codes with availability, Def.~\ref{def:LRC-t}. It is well known that in the classical
case the tradeoff between the rate and relative distance of codes on asymptotically maximal curves improves upon the Gilbert-Varshamov 
(GV) bound \cite{tvz82}. Here we point out similar improvements for the LRC versions of the GV bound.

\subsection{Asymptotic GV-type bounds for LRC codes}


To introduce the asymptotic parameters, consider the sequence of LRC codes $C_i,i=1,2,\dots$ of length $n_i$, dimension $k_i$ and distance $d_i$. We will assume that the locality parameter is fixed and equals $r.$
Suppose that $n_i\to\infty, i=1,2\dots,$ and that there exist limits $R=\lim_{i\to\infty} k_i/n_i$ and $\delta=\lim_{i\to\infty}
d_i/n_i.$ In this case we say that the code sequence $(C_i)$ has asymptotic parameters $(R,\delta).$

A bound analogous to the GV bound in the LRC case has been recently derived in 
\cite{TBF,Cad13}. 
\begin{theorem}\label{thm:GV} There exists a sequence of $q$-ary linear $r$-LRC codes with the asymptotic parameters $(R,\delta)$ 
as long as
\begin{equation}\label{eq:GV}
    R< \frac{r}{r+1} - \min\limits_{0<s\le 1} \Big\{ \frac{1}{r+1}\log_q b(s) - \delta \log_q s \Big\},
    \end{equation}
 where
 \begin{equation}\label{eq0:fg}
    b_2(s)  = \frac{1}{q}( ( 1 + (q-1)s)^{r+1} + (q-1) (1-s)^{r+1}).
    \end{equation}
    \end{theorem}

Turning to $(n,k,r,\rho)$ LRC codes with $\rho\ge 2$, i.e., codes that correct multiple erasures, we establish the
following result.
%

\vspace*{.1in}\begin{theorem}   Assume that there exists a $q$-ary MDS code of length $r+\rho-1$ and distance $\rho.$ 
For any pair of values $(R,\delta)$ such that 
  \begin{equation}\label{eq:GVrho}
   R=R_q(r,\rho,\delta) <\frac{r}{r+\rho-1}-\min_{0< s\le 1}\Big\{\frac {\log_q b_\rho(s)}{r+\rho-1}-\delta \log_q s\Big\}   
    \end{equation}
    where
    \begin{equation}  \label{eq:fgrho}
    b_\rho(s)=1+(q-1)\sum_{w=\rho}^{(r+\rho-1)}\binom {r+\rho-1}w 
       s^w q^{w-\rho}\sum_{j=0}^{w-\rho} \binom{w-1}j(-q)^{-j}.
    \end{equation}
there exists a sequence of $q$-ary linear $r$-LRC codes with the asymptotic parameters $(R,\delta)$ that correct locally $\rho-1\ge 1$ erasures.
 \end{theorem}
\vspace*{.1in}\begin{IEEEproof} (outline): The proof is a minor modification of Theorem B, Eq.(19) in \cite{TBF}, so we only outline it here.
Let $\cC$ be a linear $(n,k,r,\rho)$ LRC code over $\ff_q.$
Suppose that
$n$ is divisible by $r+\rho-1.$
Consider an $(n-k)\times n$ matrix over $\ff_q$ of the form $H=\begin{bmatrix}H_U\\H_{L}\end{bmatrix}$ where $H_U$ is a block-diagonal
matrix and $H_L$ is a random uniform matrix over $\ff_q.$ Assume that $H_U$ has the form
 \begin{equation}\label{eq:matrix}
 H_U=\begin{bmatrix}
 \text{\framebox[.4in][c]{$H_0$}}&&&\\
 &\text{\framebox[.4in][c]{$H_0$}}&&\\
 &&\ddots&\\
 &&&\text{\framebox[.4in][c]{$H_0$}}
 \end{bmatrix}
 \end{equation}
 where $H_0$ is the parity-check matrix of an $[r+\rho-1,r]$ MDS code.
This defines an ensemble of $(n-k)\times n$ matrices $\cH_q(n,k,r).$ 
We estimate the probability that the code with the parity-check matrix $H\in\cH_q$ contains a nonzero vector $x$ of weight $<d.$
First, we estimate the weight distribution of the code $\cC_U=\ker(H_U)$ using the weight enumerator $b_\rho(s)$ of
the MDS code. This gives for the number of vectors of weight $w$ in $\cC_U$ the estimate
   $$
   B_w\le \min_{0<s\le 1}{s^{-w}}{b_\rho(s)^\frac{n}{r+\rho-1}}.
   $$
   Then we use the union bound to estimate the probability that at least one of these vectors satisfies $H_Lx^T=0.$ 
   Equation \eqref{eq:GVrho} gives a sufficient condition for this probability to go to zero as $n\to\infty.$
\end{IEEEproof}

   Note that for $\rho=2$ the bound \eqref{eq:GVrho}-\eqref{eq:fgrho} reduces to \eqref{eq:GV}-\eqref{eq0:fg}.

\vspace*{.1in}
A GV-type bound for $(n,k,\{r,r\})$ codes with two recovery sets was also established in \cite{TBF}, Theorem B, Eq.(20).
For comparison with codes in this paper we would need to modify that proof to account for different sizes of the two 
recovery sets, say $r_1$ and $r_2.$ This modification is readily obtained as follows. To prove a GV-type bound in \cite{TBF}
we take local codes constructed on complete graphs on $r+2$ vertices (i.e., the matrix $H_0$ in \eqref{eq:matrix} is an edge-vertex incidence matrix
of $K_{r+2}$ from which one row is deleted to obtain a full-rank matrix). To obtain a bound for $(n,k,\{r_1.r_2\})$ LRC codes
we replace in this argument $K_{r+2}$ with the edge-vertex adjacency matrix of a complete bipartite graph $K_{r_1+1,r_2+1}$ and
follow the steps of the above proofs. The result is too cumbersome (and not too instructive) to be included in this text.

\subsection{Asymptotic parameters of codes on Garcia-Stichtenoth curves}

Let us compute the asymptotic relation between the parameters of LRC codes on the 
Garcia-Stichtenoth curves constructed above. 

\subsubsection{LRC codes correcting one erasure}
Propositions \ref{prop:GS1}-\ref{prop:GS2} lead to the following asymptotic results.
\begin{proposition} \label{prop:ab}
Let $q=q_0^2,$ where $q_0$ is a power of a prime. There exist families of LRC codes with locality $r$
whose rate and relative distance satisfy
   \begin{align}
     R&\ge \frac r{r+1}\Big(1-\delta-\frac 3{\sqrt q+1}\Big), \qquad r=\sqrt q-1   
    \label{eq:ab1}\\
      R&\ge \frac r{r+1}\Big(1-\delta-\frac{2\sqrt q}{q-1}\Big),  \qquad r=\sqrt q . \label{eq:ab2}
     \end{align}
 \end{proposition}
{\em Remark 4.4:} Recall that without the locality constraint the relation between $R$ and $\delta$ for codes on asymptotically
optimal curves (for instance, on the curves $X_l,l=2,3,\dots$) takes the form 
  $
  R \ge 1-\delta-\frac{1}{\sqrt q-1};
  $ see \cite[p.251]{TVN07}.
 \begin{IEEEproof}
For instance, let us check \eqref{eq:ab1}. From \eqref{eq:GS1} we obtain
  \begin{align}
  d+\frac{k(r+1)}{r}&\ge n_l-\frac{q_0n_{l-1}}{q_0-1}-\frac{2n_l(q_0-2)}{q_0^2-1}+q_0\nonumber\\
  &\ge n_l\Big(1-\frac{1}{q_0-1}-\frac{2q_0-4}{q_0^2-1}\Big) \label{eq:sggs}\\
  &=n_l\Big(1-\frac{3}{q_0+1}\Big).\nonumber
  \end{align}
Letting $\delta=d/n_l, R=k/n_l, l\to\infty,$ we obtain \eqref{eq:ab1}.
\end{IEEEproof}

The bound given in \eqref{eq:ab2} (i.e., the code family constructed in Prop.~\ref{prop:GS2}) improves upon the GV-type bound \eqref{eq:GV}-\eqref{eq0:fg} for
large alphabets. For instance, for $q_0=23$ the code rate \eqref{eq:ab2} is better than \eqref{eq:GV} for $\delta\in[0.413,0.711],$
and the length of this interval increases as $q_0$ becomes greater. Similar conclusions can be made for the codes in the family of Prop.~\ref{prop:GS1}.
\addtocounter{figure}{2}
\begin{figure}[ht]
 \centering {\includegraphics[scale=.5]{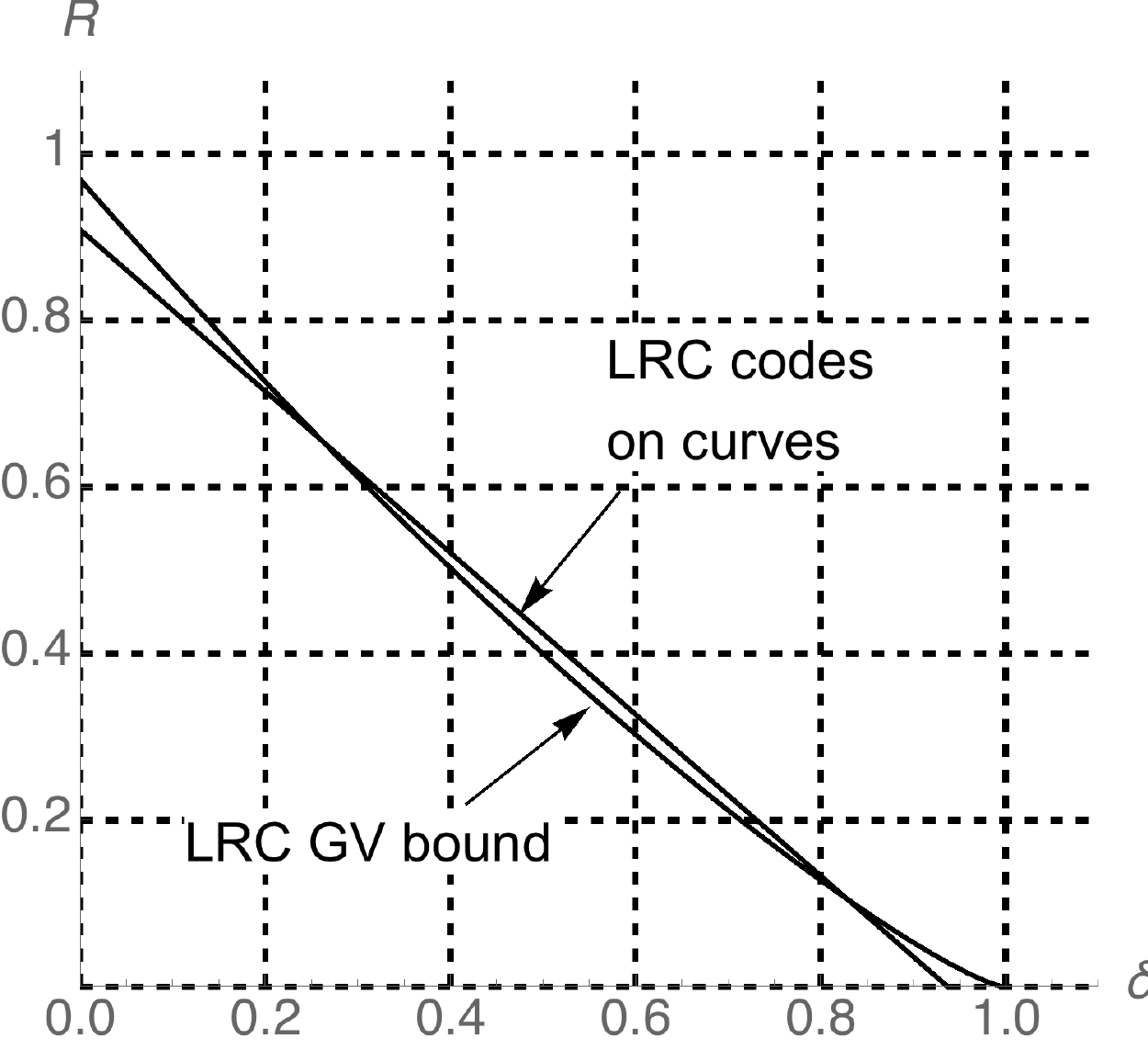}\hspace*{.5in}\includegraphics[scale=.5]{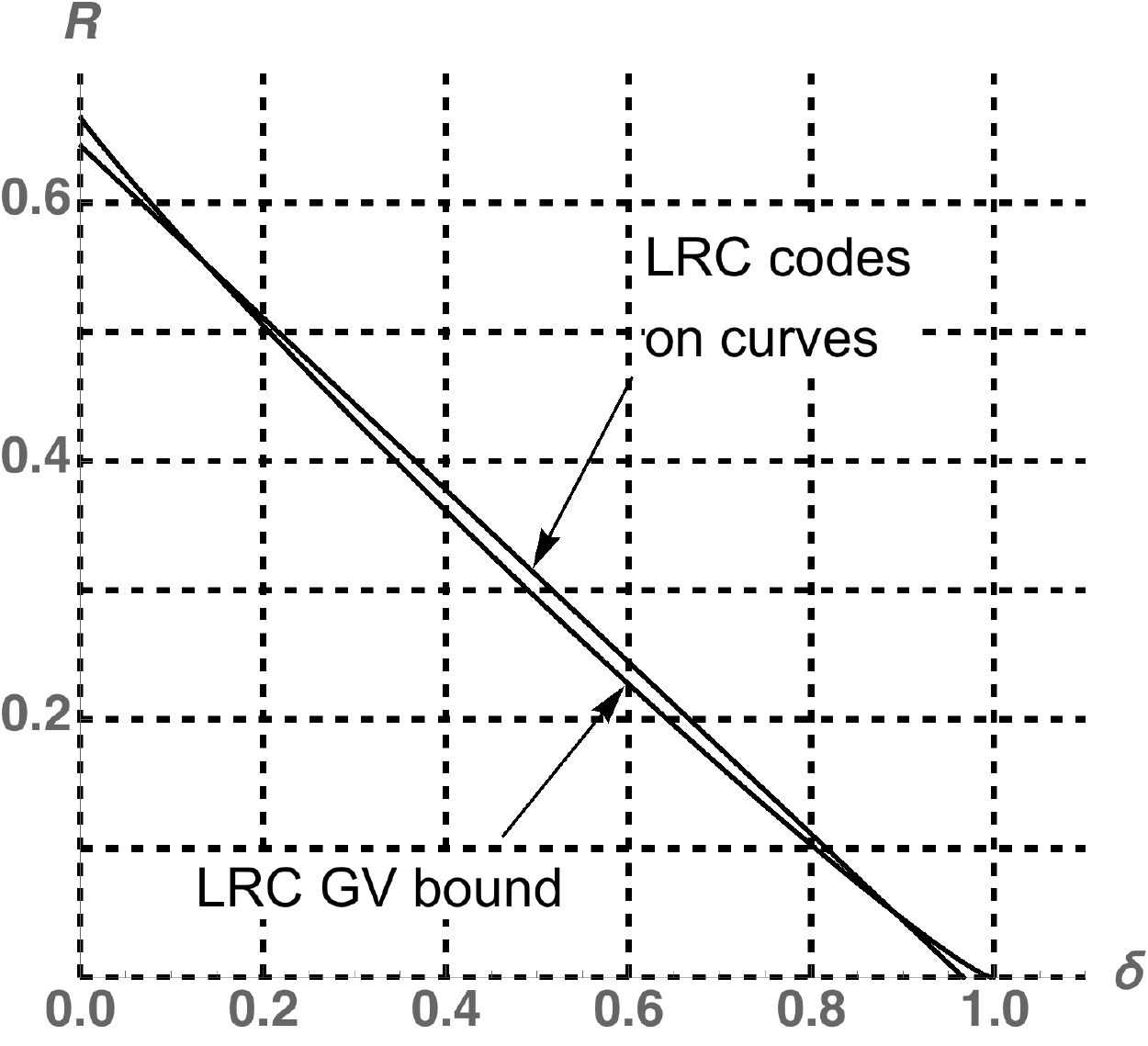}} 
   \caption[caption]{\small {\em Left plot:} The bound \eqref{eq:ab1} shown together with the Gilbert-Varshamov type bound \eqref{eq:GV} ($q_0=r=23$).
   
 \hspace*{.3in}      {\em Right plot:} The bound \eqref{eq:a-small} together with the GV-type bound, $r=2, q_0=32$.}\label{fig:ag}
  \end{figure}

Turning to codes with small locality, let us derive an asymptotic estimate of the parameters of the code family in Proposition~\ref{prop:GS4}. 
Applying the same argument as in the proof of Proposition 5.3 we see that $g_Y\le 1+\frac{G_{X_l}-1}{r+1}$ which is sufficient to prove 
Proposition 5.6 below.
In fact, one can note that the curve $Y$ is a quotient of an asymptotically optimal curve, so it is asymptotically optimal itself. Therefore,
the genus $g_Y$ is asymptotic to $G_{X_l}/(r+1)$, and the genus $ G_{X_l}$ is asymptotic to 
$q_0^{l-1}(q_0+1)=\frac{n_l}{q_0-1}$, but it is not important for the proof. 
\begin{proposition}\label{prop:small} Let $(r+1)|(q_0+1)$ and let $q=q_0^2,$ where $q_0$ is a power of a prime. There exists a family
 of $q$-ary LRC codes with locality $r$ whose 
asymptotic rate and relative distance satisfy the bound
  \begin{equation}\label{eq:small}
     R\ge \frac{r}{r+1}\Big(1-\delta-\frac{q_0+r}{q_0^2-1}\Big).
     \end{equation}
\end{proposition}
For instance, take $r=2$ and let $3|(q_0+1).$ We obtain the bound
   \begin{equation}
   R\ge \frac23\Big(1-\delta-\frac1{\sqrt{q}-1}-\frac1{q-1}\Big).
   \label{eq:a-small}
   \end{equation}
Asymptotic bounds obtained above are shown in Fig.~\ref{fig:ag} both for locality $r=q_0$ and for $r=2.$ 

\subsubsection{LRC codes correcting multiple erasures}
Now consider the case of $(n,k,r,\rho)$ LRC codes with $\rho\ge 2$. From Proposition \ref{prop:GS2-rho}
we obtain the following result.
\begin{proposition} \label{prop:asymp-rho} Let $q=q_0^2,$ where $q_0$ is a power of a prime. The rate and relative distance of LRC codes 
   \begin{equation}\label{eq:asymp-rho}
   R\ge \frac r{r+\rho-1}\Big(1-\delta-\frac 3{q_0+1}\Big)
   \end{equation}
   for any $r,\rho$ such that $r+\rho-1=q_0.$
\end{proposition}   
\textcolor{black}{
\begin{IEEEproof} We essentially repeat the calculation in \eqref{eq:sggs}. Note that in this case, according to \eqref{eq:rho},
the Singleton gap is computed in the form $d+\frac {k(r+\rho-1)}{r},$ so from \eqref{eq:GSrho1} we obtain
  $$
  d+\frac {k(r+\rho-1)}{r}\ge n_l-\frac{q_0n_{l-1}}{q_0-1}-\frac{2n_l(q_0-2)}{q_0^2-1}+q_0.
  $$
Now \eqref{eq:asymp-rho} follows from \eqref{eq:sggs} by taking the limit $l\to\infty.$
\end{IEEEproof}}

\vspace*{.1in}
To give an example, take $\rho=3,q_0=43,$ then $r=41.$ Then \eqref{eq:asymp-rho} improves
upon the GV-type bound \eqref{eq:GVrho} for $0.486\le R\le 0.685.$

\vspace*{.1in}\subsubsection{LRC codes with two recovery sets} Finally, consider codes with the availability property. Letting in Proposition
\ref{prop:GS2a} $n=n_l\to\infty,$ we obtain the following statement.
\vspace*{.05in}
\begin{proposition} Let $q=q_0^2,$ where $q_0$ is a power of a prime, and suppose that $(r_1+1)|(q_0+1)$ and $(r_2+1)|q_0$.
There exits a family of $q$-ary $(n,k,\{r_1,r_2\})$ LRC codes whose asymptotic rate $R$ and relative distance $\delta$
satisfy the relation
  \begin{equation}\label{eq:2}
  \delta+\frac{(r_1+1)(r_2+1)}{r_1r_2}R\ge\frac{q_0-2}{q_0-1}-\frac{(r_1+r_2-2)}{q_0^2-1}. 
  \end{equation}
  \end{proposition}
In the case of a single recovery set we evaluated the quality of our constructions by computing the Singleton gap (see, e.g., \eqref{eq:sg}, or the proofs of Propositions \ref{prop:ab}, \ref{prop:asymp-rho}). From the Singleton-like bound \eqref{eq:tbf} we obtain the relation
  $$
  \delta+\frac{r^2+r+1}{r^2}R\ge 1.
  $$
At the same time, assuming (with no justification) that $r_1=r_2$ in \eqref{eq:2}, 
we obtain the relation
   $$
   \delta+\frac{r^2+2r+1}{r^2}R\ge \frac{q_0-2}{q_0-1}-2\frac{r-1}{q-1}
   $$
which does not differ from the Singleton bound by much for large $q_0.$

\providecommand{\bysame}{\leavevmode\hbox to3em{\hrulefill}\thinspace}
\providecommand{\MR}{\relax\ifhmode\unskip\space\fi MR }
\providecommand{\MRhref}[2]{%
  \href{http://www.ams.org/mathscinet-getitem?mr=#1}{#2}
}
\providecommand{\href}[2]{#2}

\end{document}